\documentclass[a4paper,11pt]{article}
\usepackage{pos}
\usepackage{bibspacing}

\title{Higgs production in the high-energy limit of pQCD}

\author[a,b,c]{F. G. Celiberto}
\author*[d,e]{M. Fucilla}
\author[f]{D. Yu. Ivanov}
\author[d,e]{M.M.A. Mohammed}
\author[d,e]{A. Papa}

\affiliation[a]{European Centre for Theoretical Studies in Nuclear Physics and Related Areas (ECT*),
I-38123 Villazzano, Trento, Italy}

\affiliation[b]{Fondazione Bruno Kessler (FBK), I-38123 Povo, Trento, Italy}

\affiliation[c]{INFN-TIFPA Trento Institute of Fundamental Physics and Applications,
I-38123 Povo, Trento, Italy}

\affiliation[d]{Dipartimento di Fisica, Università della Calabria,
I-87036 Arcavacata di Rende, Cosenza, Italy}

\affiliation[e]{Istituto Nazionale di Fisica Nucleare, Gruppo collegato di Cosenza,
I-87036 Arcavacata di Rende, Cosenza, Italy}

\affiliation[f]{Sobolev Institute of Mathematics, 630090 Novosibirsk, Russia}

\emailAdd{francescogiovanni.celiberto@unipv.it}
\emailAdd{michael.fucilla@unical.it}
\emailAdd{d-ivanov@math.nsc.ru}
\emailAdd{mohammed.maher@unical.it}
\emailAdd{alessandro.papa@fis.unical.it}

\abstract{With the advent of TeV-energy colliding machines, such as the Large Hadron Collider (LHC), the possibility has opened up to test predictions of Quantum Chromodynamics (QCD) and, more in general, of the Standard Model (SM), in new, and so far unexplored, kinematical regimes. Among the many reactions that can be investigated at LHC, the Higgs production is one of the most important and challenging for the entire high-energy physics Community. Beside usual studies in the Higgs sector, it has recently been highlighted how differential Higgs distributions can be effectively used as ``stabilizers'' of the high-energy dynamics of QCD. The definition and the study of observables sensitive to high-energy dynamics in Higgs production has the double advantage of (i) allowing us to clearly disentangle the high-energy dynamics from the fixed-order one and (ii) providing us with an auxiliary tool to extend Higgs studies in wider kinematical regimes. \\
In this work, we will show how a general hybrid collinear/high energy factorization can be built up for the inclusive production of a Higgs in association with a jet. Then, we will present some phenomenological analyses that corroborate the underlying assumption that this reaction can be used to investigate the semi-hard regime of QCD. Finally, we will focus on more formal developments, such as the inclusion of subleading corrections to previous studies, via the calculation of the forward next-to-leading order Higgs impact factor.}

\FullConference{%
  *** Particles and Nuclei International Conference - PANIC2021 ***\\
  *** 5 - 10 September, 2021 ***\\
  *** Online ***
}

\begin{document}
\maketitle

\section{Introduction}
The high energies in the center-of-mass reached at the Large Hadron Collider allow to investigate hadronic reactions in new kinematical regimes. A particularly interesting one is the so-called \textit{semi-hard regime}, characterized by a center-of-mass energy, $\sqrt{s}$, much larger than the hard scales of the process, $\{ Q \}$, which are, in turn, much larger than the QCD mass scale, $\Lambda_{{\rm{QCD}}}$. In this regime the usual approach based on fixed-order calculations in perturbation theory must be replaced, or at least supplemented, by a method that allows the resummation of the large energy-logarithms, entering the perturbative series with powers increasing with the perturbative order. 
The Balitsky-Fadin-Kuraev-Lipatov (BFKL) approach~\cite{Fadin:1975cb,Kuraev:1976ge,Kuraev:1977fs,Balitsky:1978ic} is a systematic method for the resummation of these large contribution, both in the leading logarithmic approximation, LLA, and in the next-to-leading logarithmic approximation, NLA. In this framework, the cross sections of processes take a peculiar factorized form, given by the convolution of two process-dependent \textit{impact factors}, portraying the transition from each colliding particle to the respective final-state object (see~\cite{Celiberto:2020wpk} for a list of known impact factors), and a process-independent \textit{Green’s function}. Processes with two identified objects separated by a large rapidity interval can be investigated via the so-called \textit{hybrid} collinear/high-energy factorization, where collinear ingredients, such as parton distribution functions (PDFs), fragmentation functions (FFs) and jet functions (JFs), enter the definiton of BFKL impact factors.

In the following, we will focus on the inclusive hadroproduction of a Higgs boson and a jet,
\begin{equation}
    {\rm{proton}}(p_1) + {\rm{proton}}(p_2) \longrightarrow {\rm{jet}} (p_J) + X + {\rm{Higgs}} (p_H)  \; ,
\end{equation}
where $X$ stands for an undetected hadronic system.

\section{Phenomenology}
We consider in our analysis two observables: 
\begin{itemize}
    \item \textit{Azimuthal correlation moment} $C_1/C_0$, where
    \begin{equation}
\label{C0}
 C_n(s, \Delta Y) = 
 \int_{y_H^{\rm min}}^{y_H^{\rm max}} d y_H
 \int_{y_J^{\rm min}}^{y_J^{\rm max}} d y_J
 \int_{p_H^{\rm min}}^{p_H^{\rm max}} d |\vec p_H|
 \int_{p_J^{\rm min}}^{p_J^{\rm max}} d |\vec p_J| \; 
 \delta (\Delta Y - (y_H - y_J)) \;
 {\cal C}_n
 \; .
\end{equation}
The complete definition of ${\cal C}_n$ can be found in~\cite{Celiberto:2020tmb}.
    \item Higgs $p_T$-\textit{distribution}
    \begin{equation}
\label{pH}
 \frac{d \sigma(|\vec p_H|, s, \Delta Y)}{d |\vec p_H| d \Delta Y} = 
 \int_{y_H^{\rm min}}^{y_H^{\rm max}} d y_H
 \int_{y_J^{\rm min}}^{y_J^{\rm max}} d y_J
 \int_{p_J^{\rm min}}^{p_J^{\rm max}} d |\vec p_J| \;
 \delta (\Delta Y - (y_H - y_J)) \;
 {\cal C}_0 \;.
\end{equation}
\end{itemize} 
For both the observables we constrain the Higgs emission inside rapidity acceptances of the CMS barrel detector, $|y_H| < 2.5$, while we allow for a larger rapidity range of the light jet, that can be detected also by the CMS endcaps, $|y_J| < 4.7$. We consider a \emph{symmetric} configuration of $p_T$-ranges: $20 < |\vec p_{J,H}|/{\rm GeV} < 60$ for $C_1/C_0$, while we set $35 < |\vec p_J|/{\rm GeV} < 60$ for the $|\vec{p_H}|$-distribution. The center-of-mass energy squared is fixed at $\sqrt{s} = 14$~TeV.

\begin{figure}
    \centering
    \includegraphics[scale=0.40]{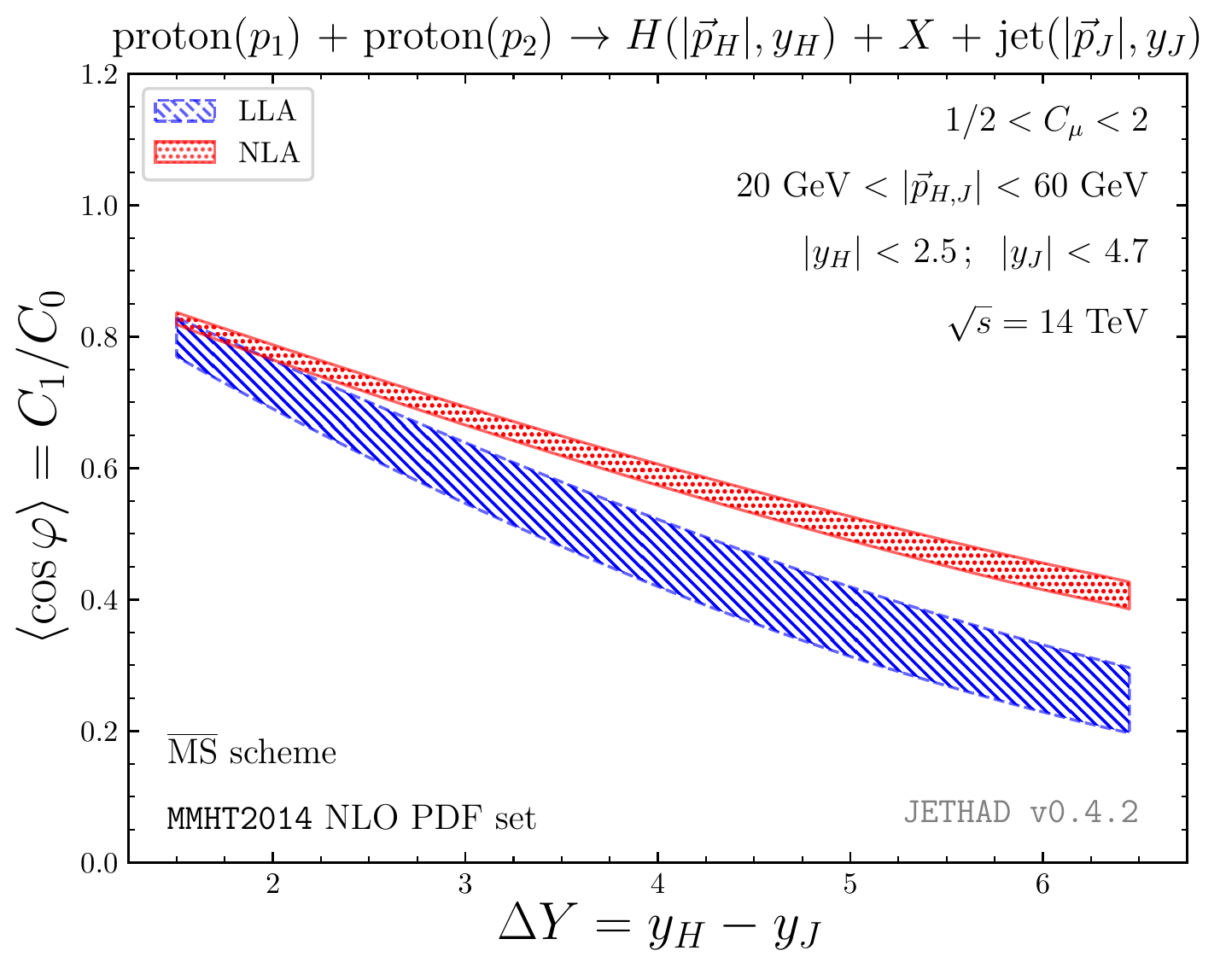}
    \includegraphics[scale=0.40]{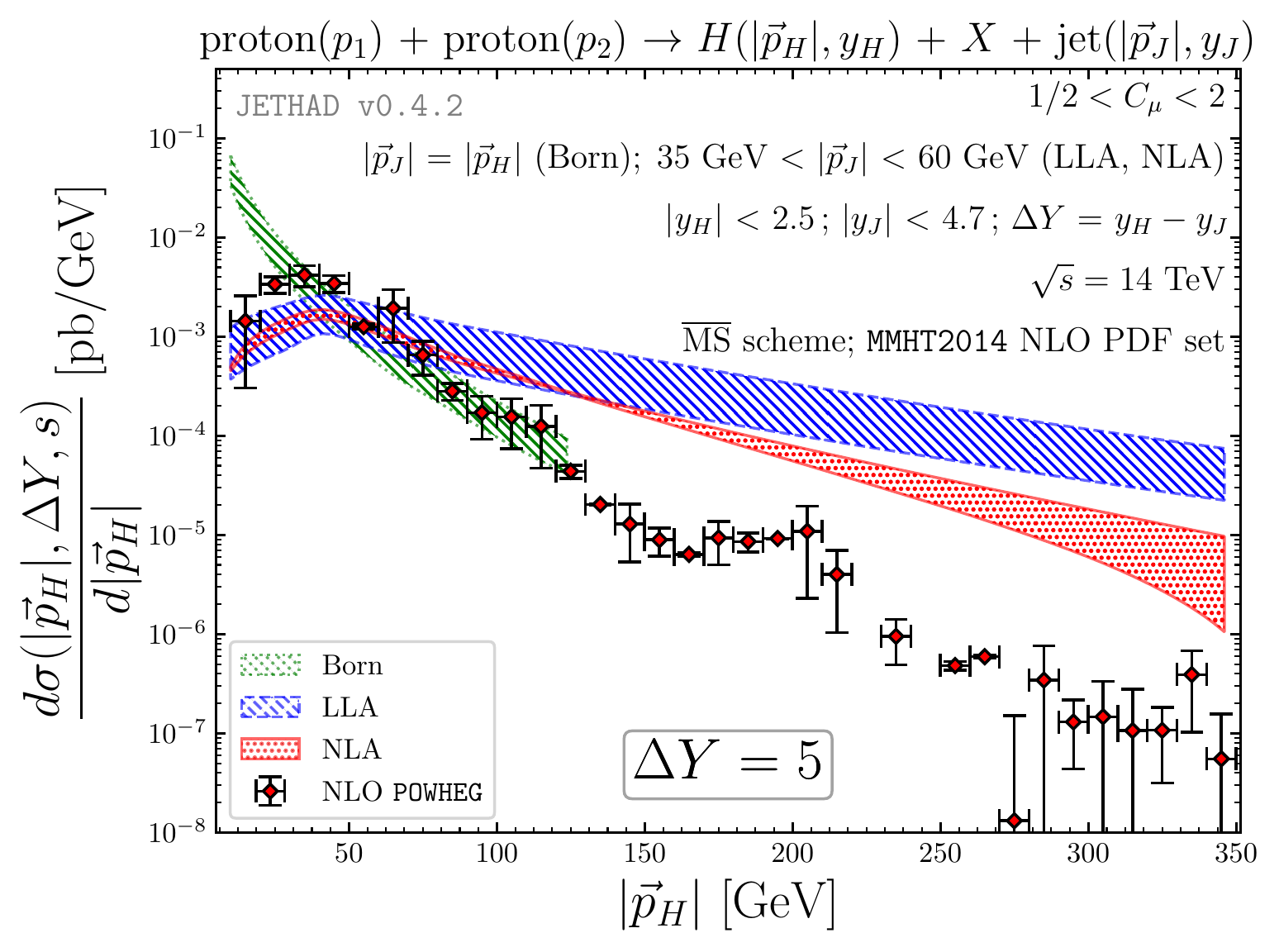} 
    \caption{$C_1/C_0$ as a function of $\Delta Y$ (left) and Higgs transverse-momentum distribution at $\Delta Y = 5$ (right). Text boxes inside panels show transverse-momentum and rapidity ranges.}
    \label{Fig:Co}
\end{figure}

 From the left panel of \ref{Fig:Co} we see that considering the production of the Higgs gives us the chance to obtain a natural "stabilizing" effect under higher-order corrections and under scale variations. This feature, absent in the production of two jets strongly separated in rapidity
(Muller-Navelet channel), allows us to carry out studies around natural scales, avoiding the implementation of optimization procedures of the series such as Brodsky-Lepage-Mackenzie one. A similiar stabilization of the series has been observed in heavy-flavor production~\cite{Bolognino:2019yls,Bolognino:2021mrc,Celiberto:2021dzy,Celiberto:2021fdp}. Furthermore, it is important to underline that, in the kinematic region in which the BFKL approach is suitable for describing the reaction ($ \vec{p} _H \sim \vec{p}_J $), the Higgs $p_T$-distribution is stable under higher-order corrections and its trend deviates from the fixed order NLO prediction obtained through POWHEG method \cite{Frixione:2007vw,Alioli:2010xd}.

\section{Towards a next-to-leading description}

Our current goal is the inclusion of subleading effects due to next-to-leading order corrections to the Higgs impact factor.  We consider the next-to-leading order impact factor in the infinite top-mass limit 
through the effective Lagrangian
\begin{equation}
\mathcal{L} = \frac{1}{4} g_H F_{\mu \nu}^{a} F^{\mu \nu, a}H
\label{lagrangian}
\end{equation}
This will generate vertices in which two, three or four gluons couple directly with the Higgs. 
The LO order impact factor is very simple and reads
\begin{equation}
\frac{d \Phi_{gg}^{H} (z_H, \vec{p}_H, \vec{q})}{d z_H d^2 \vec{p}_H} = \frac{g_H^2}{8 \sqrt{N_c^2-1}} \vec{q}^{\; 2} \delta (1-z_H) \delta^{(2)} (\vec{q}-\vec{p}_H) \; ,
\end{equation}
where $z_H$ is the fraction of momenta of the initial gluon carried by the Higgs, $\vec{p}_H$ its transverse momenta and $\vec{q}$ is the Reggeon transverse momenta. The basic ingrediants to build the NLO impact factor are: the real contribution associated with the emission of an additional quark ($d \Phi_{q q}^{ \{H q \}}$), the real contribution associated with the emission of an additional gluon ($d \Phi_{g g}^{ \{H g \}}$), the virtual contribution ($d \Phi_{g g}^{ \{H \} (\rm{1-loop})}$). These contributions are combined with a suitable "BFKL" counter term to remove \textit{rapidity divergences} and then convoluted with corresponding PDFs to reabsorb \textit{infrared singularities} affecting partonic impact factors. The remaining \textit{soft singularities} cancel out when we combine real and virtual corrections, as guaranteed by the Kinoshita-Lee-Nauenberg theorem. At this point, we are left with only \textit{ultraviolet divergences} which are removed by the renormalization procedure.

The inclusion of these effects has already been investigated, in the case of single forward Higgs production~\cite{Hentschinski:2020tbi,Nefedov:2019mrg}, and, by using the Lipatov effective action~\cite{Lipatov:1995pn}. Hence, it will be also interesting to check compatibility and consistency between the two approaches. 

\section{Conclusion and outlook}
A high-energy treatment can
be afforded in Higgs plus jet hadroproduction in the region in which $|\vec{p}_H| \sim |\vec{p}_J|$ and it exhibits quite a fair stability under higher-order corrections. The definition and the study of observables sensitive to high-energy dynamics in Higgs production has the double advantage of:
\begin{itemize}
    \item allowing us to clearly disentangle the high-energy dynamics from the fixed-order one.
    \item providing us with an auxiliary tool to extend Higgs studies in wider kinematical regimes.
\end{itemize}
We are currently upgrading our study to next-to-leading order in the limit $m_t \rightarrow \infty$. From the phenomenological point of view, in future dedicated studies, it could be interesting to investigate the single Higgs production (both in forward and central region of rapidity) by introducing the \textit{unintegrated gluon distribution} (UGD). The inclusion of top-mass effects can also be an interesting development.

\vspace{-0.25cm}
\bibliographystyle{apsrev}
\bibliography{references}

\end{document}